\documentclass[aps,pre,twocolumn,showpacs,superscriptaddress,10pt]{revtex4}
\usepackage{graphicx}
\usepackage{subfigure}
\usepackage{amsmath}
\usepackage{amsfonts}
\usepackage{amssymb}
\usepackage{amsthm}
\usepackage{times}
\usepackage[colorlinks,citecolor=blue,linkcolor=blue]{hyperref}
\usepackage{color}

\begin{document}

\title{Variational approach to renormalized phonon in momentum-nonconserving nonlinear lattices}

\newcommand{\Fudan}{\affiliation{State Key Laboratory of Surface
Physics and Department of Physics, Fudan University, Shanghai 200433,
China}}
\newcommand{\Colorado}{\affiliation{Department of Mechanical Engineering, University of Colorado, Boulder, CO 80309,
USA}}
\newcommand{\Nanjing}{\affiliation{Collaborative Innovation Center of Advanced Microstructures, Fudan University, Shanghai 200433, China}}

\author{Junjie Liu}
\Fudan
\author{Baowen Li}
%\email{phylibw@nus.edu.sg}
\Colorado
\author{Changqin Wu}
%\email{cqw@fudan.edu.cn}
\Fudan
\Nanjing

\newcommand{\LH}{{LH}}
\newcommand{\UH}{{UH}}

\begin{abstract}
A previously proposed variational approach for momentum-conserving systems [J. Liu et.al., Phys. Rev. E {\bf91}, 042910 (2015)] is extended to systematically investigate general momentum-nonconserving nonlinear lattices. Two intrinsic identities characterizing optimal reference systems are revealed, which enables us to derive explicit expressions for optimal variational parameters. The resulting optimal harmonic reference systems provide information for the band gap as well as the dispersion of renormalized phonons in nonlinear lattices. As a demonstration, we consider the one-dimensional $\phi^4$ lattice. By combining the transfer integral operator method, we show that the phonon band gap endows a simple power-law temperature dependence in the weak stochasticity regime where predicted dispersion is reliable by comparing with numerical results. In addition, an exact relation between ensemble averages of the $\phi^4$ lattice in the whole temperature range is found, regardless of the existence of the strong stochasticity threshold.
\end{abstract}

\date{}

%63.20.-e	Phonons in crystal lattices
%45.10.Db	Variational and optimization methods
%63.20.Ry	Anharmonic lattice modes
%05.45.-a	Nonlinear dynamics and chaos

\maketitle

\section{Introduction}
The understanding of the microscopic foundation of Fourier's law is a longstanding problem in statistical physics. One of the promising ways along this direction is the study of heat conduction in low dimensional nonlinear lattices \cite{Lepri.03.PR,Dhar.08.AP,Liu.12.EPJB}. Among the investigations, an interesting perspective is to find a nonlinear lattice which displays normal heat conduction, namely, where Fourier's law holds. After numerous numerical simulations \cite{Casati.84.PRL,Gillan.85.JPC,Prosen.92.JPA,Hu.98.PRE,Hu.00.PRE,Aoki.00.PLA}, a consensus that one-dimensional (1D) nonlinear lattices with on-site potentials can establish normal heat conduction has been reached. Unraveling the microscopic mechanism of heat transfer across such momentum-nonconserving (MNC) lattices thus may provide crucial insights to the foundation of Fourier's law.

Phenomenologically \cite{Peierls.55.NULL}, the heat conductivity $\kappa$ can be approximately obtained via the Peierls formula, i.e., $\kappa=\sum_kC_kv_kl_k$, here $k$ is the phonon wave number, $C_k$ the specific heat of the phonon mode, $v_k$ its phonon group velocity and $l_k$ the corresponding mean free path (MFP). Following this spirit, a theoretical proposal for phonon heat transport based on renormalized phonon has been carried out \cite{Li.07.EL,Li.13.PRE}. Nevertheless, flaws in its phonon description \cite{Zhang.13.A} may produce wrong phonon information, which limits its ability and generality. Trying to understand the process of heat conduction from the phonon point of view, theoretical investigations on the phonon properties seems to be a necessary step. Thus a unified phonon theory which is capable of dealing with various nonlinear lattices is of primary interest and importance at present stage. However, the attempts concerning phonon properties in MNC lattices are less \cite{Li.07.EL,Li.13.PRE,Yang.14.PRE}, compared with studies carried on momentum-conserving (MC) counterparts exhibiting anomalous heat conduction
\cite{Alabiso.95.JSP,Lepri.98.PRE,Alabiso.01.JPA,Gershgorin.05.PRL,Li.06.EL,Gershgorin.07.PRE,He.08.PRE,Li.10.PRL,Liu.15.PRE}.
More efforts should be devoted to the former.

In a previous study \cite{Liu.15.PRE}, we have demonstrated the advantages of a variational approach in describing renormalized phonons in MC nonlinear lattices with either symmetric or asymmetric nearest-neighbor potentials compared with existing quasi-harmonic theories. Moreover, besides our method, we found that the nonlinear fluctuating hydrodynamics (NFH) \cite{Spohn.15.JSP} can also provide good estimation for the sound velocity of MC nonlinear lattices. Since the concept of sound velocity is invalid in MNC nonlinear lattices, we deduce that the NFH can not be able to give information of the renormalized phonons, although it can still describe correlation functions of these systems \cite{Spohn.14.A}. Therefore, an interesting question arises whether the variational approach can be exploited for this class of models.

The purpose of this work is thus twofold: (i)By putting the proposed variational approach onto a more general ground, we show that MNC nonlinear lattices can be tackled within the framework of the variational approach, which makes our theory a candidate for a unified phonon theory, and (ii) we determine temperature-dependent phonon properties of MNC nonlinear lattices using this method and confirm the validity of so-obtained results with the aids of molecule dynamics (MD) simulations as well as a tuning fork method \cite{Liu.14.PRB}. As a demonstration, we consider one of archetype 1D nonlinear lattices in classical statistical mechanics, namely, 1D $\phi^4$ lattice. We conclude that the approach opens a new perspective to understand renormalized phonons in various nonlinear lattices.

The paper is organized as follows. In Sec. II, we present the general aspects of the extended variational approach. Two intrinsic relation satisfied by the optimal reference systems are revealed such that general nonlinear lattice can be tackled. Explicit forms of optimal variational parameters are also obtained. In Sec.III, we focus on a comprehensive study of a specific nonlinear lattice, namely, the 1D $\phi^4$ lattice. We found a power-law temperature dependence of the phonon band gap valid in the weak stochasticity regime as well as a rigorous relation between ensemble averages of the nonlinear lattice. Finally, we briefly summarize our main findings in Sec.IV.

\section{The methodologies}
We consider 1D nonlinear lattices described by the
general Hamiltonian \cite{Li.12.RMP}
\begin{equation}\label{eq:H}
H~=~\sum_{n=1}^{N}\left[\frac{p_n^2}{2}+V(q_n-q_{n-1})+ U(q_n)\right],
\end{equation}
where $N$ is the particle number, $p_n$ denotes the momentum of $n$-th particle, $q_n=x_n- n a$
denotes the displacement of $n$-th particle from its equilibrium position $na$ with $x_n$ the absolute position and $a$ the equilibrium distance for
the interaction bond (for systems with zero pressure, it coincides with the lattice spacing),
and $V$ represents the interparticle potential, $U$ is the on-site potential. We use periodic boundary conditions such that $q_0=q_N$. For brevity and without loss of generality, we take $m=1$ and $a=1$ as the unit of mass and length, respectively.

For simplicity, we introduce $\delta_n\equiv q_n-q_{n-1}$. The system evolves according to the equations of motion (EOMs)
\begin{eqnarray}
\dot{q}_n &=& p_n,\\
\dot{p}_n &=& V'(\delta_{n+1})-V'(\delta_n)-U'(q_n),
\end{eqnarray}
where the dot and the prime denote the time and spatial derivative, respectively. Notice that lattices with asymmetric interparticle potentials ($V(\delta)\neq V(-\delta)$) will induce nonvanishing internal pressure, here we only consider lattices with symmetric ones. The extension to systems involving asymmetric potentials is straightforward \cite{Liu.15.PRE}.  Accordingly, nonlinear potentials with the following forms are of special interest:
\begin{eqnarray}
V(\delta_n) &\equiv & \sum_{s=1}^{\infty}\frac{f_{2s}}{2s}\delta_n^{2s},\label{eq:v_n}\\
U(q_n)&\equiv & \sum_{k=1}^{\infty}\frac{g_{2k}}{2k}q_n^{2k}.\label{eq:u_n}
\end{eqnarray}
Coefficients are chosen in such a way that potentials exhibit a single-well character.

In order to utilize a variational approach, the basic routine is to introduce a reference system with a trial Hamiltonian $H_0$ and prepare the nonlinear system and the reference system
at the same temperature, then determines optimal reference systems from variational principles \cite{Girardeau.07.NULL}, the resulting optimal variational parameters inevitably take temperature dependence.

For systems with zero pressure, the Helmholtz free energy $F$ determines its thermodynamic properties. Since nonlinear lattices in thermal equilibrium behave like weakly interacting renormalized phonons \cite{Gershgorin.05.PRL,Gershgorin.07.PRE,Liu.14.PRB}, the free energy of the nonlinear lattices can be divided into two parts, one is the contribution of the free renormalized phonon gas described by a harmonic Hamiltonian, the other comes from the phonon-phonon interaction. However, the interaction is complicated, approximations should be taken. Intuitively, we adopt the well-known first-order cumulant inequalities of the free energy \cite{Feynman.82.NULL}
\begin{equation}
\label{eq:up}
F~\leq~F_0+\langle H-H_0\rangle_{\rho_0},
\end{equation}
and
\begin{equation}
\label{eq:low}
F~\geq~F_0+\langle H-H_0\rangle_{\rho}
\end{equation}
as the variational principles, where $F_0$ is the Helmholtz free energy of the reference system, $\rho_0 [\equiv e^{-\beta_T (H_0-F_0)}]$ and $\rho [\equiv e^{-\beta_T (H-F)}]$ stand for the canonical measure of the reference system and the nonlinear lattice, respectively, $\beta_T\equiv 1/T$ is the inverse temperature (we set $k_B=1$).

The trial Hamiltonian $H_0$ contains a set of adjustable parameters $\{\chi_i\}$, an optimal reference system can be obtained by varying those parameters such that bounds of the variational principle go to a relative minimum or maximum. Generally, without specific forms of $H$ and $H_0$, the condition for the bounds [c.f., Eqs. (\ref{eq:up}) and (\ref{eq:low})] to be stationary with respect to one of the parameters $\chi_i$ is equivalent to
\begin{equation}\label{eq:general_up}
\left\langle (H-H_0);\frac{\partial H_0}{\partial \chi_i}\right\rangle_{\rho_0}~=~0,
\end{equation}
and
\begin{equation}\label{eq:general_low}
\left\langle\frac{\partial H_0}{\partial \chi_i}\right\rangle_{\rho_0}~=~\left\langle \frac{\partial H_0}{\partial \chi_i}\right\rangle_{\rho},
\end{equation}
for  Eqs. (\ref{eq:up}) and (\ref{eq:low}), respectively, where $\langle A; B\rangle$ stands for the connected expectation value (cumulant): $\langle AB\rangle-\langle A\rangle\langle B\rangle$. These two identities result from intrinsic properties of the variational principles. Only those reference systems who satisfy one of the two identities can be regarded as optimal reference systems. For later convenience, we refer the optimal system with the property Eq. (\ref{eq:general_up}) or Eq. (\ref{eq:general_low}) to the upper bound harmonic (\UH{}) or lower bound harmonic (\LH{}) system, respectively.

The main concern of the present work is to obtain the phonon information in the nonlinear lattice from a theoretical point of view. Phonon bears a solid basis only in harmonic systems, therefore, a harmonic reference system is chosen in the theory. Anharmonic references such as the Toda systems \cite{Ferguson.82.JCP} are out of the present scope. For homogeneous nonlinear lattices with the chosen nearest-neighbor interactions as well as on-site potentials [c.f., Eqs. (\ref{eq:v_n}) and (\ref{eq:u_n})], it is sufficient to consider the following trial harmonic Hamiltonian \cite{Dauxois.93.PRE}
\begin{equation}\label{eq:h0}
H_0~=~\sum_{n=1}^{N}\left[\frac{p_n^2}{2}+\frac{\Omega^2}{2}\delta_n^2+ \frac{\gamma}{2}q_n^2\right],
\end{equation}
in which $\Omega^2$, and $\gamma$ are variational parameters with $\Omega^2$ being the effective elastic constant, and $\gamma$ the strength of the on-site potential. Eqs. (\ref{eq:general_up}) and (\ref{eq:general_low}) determine their optimal values for the \UH{} and \LH{} system, respectively (see the following). The corresponding dispersion relation reads
\begin{equation}\label{eq:h0d}
\omega_k^2~=~4\Omega^2\sin^2\frac{k a}{2}+\gamma
\end{equation}
with $k$ is the wavenumber and $a$ the lattice spacing. As can be seen, $\gamma$ also quantifies the phonon band gap. This dispersion relation with optimal parameters can be regarded as estimations of the actual dispersion in nonlinear lattices.

With the parameter set $\{\chi_i\}$ now being $\{\Omega^2, \gamma\}$, using the identities Eqs. (\ref{eq:general_up}) and (\ref{eq:general_low}), we can obtain simple expressions for the corresponding optimal parameters (see the details in Appendix \ref{sec:opt_para}). For the \UH{} system, we will get the following self-consistent equations
\begin{eqnarray}
\Omega_U^2 &=& \frac{\sum\limits_{s=1}f_{2s}\left\langle \delta_n^{2s}\right\rangle_{\rho_0}}{\left\langle \delta_n^2\right\rangle_{\rho_0}},\label{eq:omega_up}\\
\gamma_{U} &=& \frac{\sum\limits_{k=1}g_{2k}\left\langle q_n^{2k}\right\rangle_{\rho_0}}{\left\langle q_n^2\right\rangle_{\rho_0}}.\label{eq:gamma_up}
\end{eqnarray}
With $\Omega^2_U$ and $\gamma_U$, estimations of the phonon dispersion [Eq. (\ref{eq:h0d})] as well as phonon band gap via the \UH{} system are thus determined.

While for the \LH{} system, we have similar forms as follows
\begin{eqnarray}
\Omega_L^2 &=& \frac{\sum\limits_{s=1}f_{2s}\left\langle \delta_n^{2s}\right\rangle_{\rho}}{\left\langle \delta_n^2\right\rangle_{\rho}},\label{eq:omega_low}\\
\gamma_{L} &=& \frac{\sum\limits_{k=1}g_{2k}\left\langle q_n^{2k}\right\rangle_{\rho}}{\left\langle q_n^2\right\rangle_{\rho}}.\label{eq:gamma_low}
\end{eqnarray}
However, unlike $\Omega^2_U$ and $\gamma_U$, these two parameters are totally determined by ensemble averages of the nonlinear lattice. With $\Omega^2_L$ and $\gamma_L$, we can find another estimation of the phonon band gap and the phonon dispersion. It is worthwhile to mention that the above results can be applied to MC lattices, actually, $\Omega_U^2$ and $\Omega_L^2$ reduce to the known results for the FPU-$\beta$ lattice \cite{Liu.15.PRE}.

\section{The $\phi^4$ lattice}
To verify the predictions of the variational approach, we consider
the 1D $\phi^4$ lattice described by the following potentials \cite{Hu.98.PRE,Hu.00.PRE,Aoki.00.PLA}
\begin{equation}
V(\delta)=\frac{K}{2}\delta^2,~~U(q)=\frac{\lambda}{4}q^4,
\end{equation}
which corresponds to a special case of Eqs. (\ref{eq:v_n}) and (\ref{eq:u_n}) with $f_2=K$ and $g_4=\lambda$, respectively, and all other terms vanish. It is evident that only the on-site potential exhibits a nonlinearity. Being one of archetype 1D nonlinear
lattices in classical statistical mechanics, this MNC lattice has two advantages, on the one hand, the existence of renormalized phonons in this model has been confirmed by the use of the tuning fork method very recently \cite{Liu.14.PRB}, thus we can compare theoretical predictions with numerical results, on the other hand, the dynamics of this model has been studied extensively, especially the existence of a strong stochasticity threshold (SST) which separates different phase-space structure as well as dynamic behaviors \cite{Pettini.90.PRA,Pettini.91.PRA}. It turns out that the temperature dependence of phonon band gap will manifest such transition behavior. The SST also determines the validity regime of the optimal harmonic reference systems.
In the following, we will give a detailed study of the $\phi^4$ lattice by using our variational approach together with numerical simulations. For simplicity and without loss of generality, we only present details with $K=1$ and $\lambda=1$, results with other parameter sets share similarities.

\subsection{Predictions of the variational approach}
Before going into details, we observe that $\Omega_U^2$ and $\Omega_L^2$ [c.f., Eqs. (\ref{eq:omega_up}) and (\ref{eq:omega_low})] should equal $1$ by noting that the interparticle potential of the 1D $\phi^4$ lattice is quadratic. Thus only $\gamma$ needs further efforts.

Firstly, we discuss the result obtained in the \UH{} system. The numerator of Eq. (\ref{eq:gamma_up}) consists of only one term $\langle q_n^4\rangle_{\rho_0}$. For a quadratic Hamiltonian, it is straightforward to show that $\langle q_n^4\rangle_{\rho_0}=3\langle q_n^2\rangle_{\rho_0}^2$ \cite{Dauxois.93.PRE}, then the self-consistent equation [Eq. (\ref{eq:gamma_up})] reduces to
\begin{equation}
\gamma_U~=~3\langle q_n^2\rangle_{\rho_0}.
\end{equation}
Note that $\langle q_n^2\rangle_{\rho_0}$ can be directly computed from the Helmholtz free energy $F_0$ via
\begin{equation}\label{eq:ff}
 \sum\limits_n\langle q_n^2\rangle_{\rho_0}~=~2\frac{\partial}{\partial\gamma}F_0.
\end{equation}
For the harmonic system, $F_0$ can be expressed as $-\beta_T^{-1}\sum_k\ln\frac{2\pi }{\beta_T\omega_k}$ with $\omega^2_k=4 \sin^2\frac{ka}{2}+\gamma$, the above equation thus leads to
\begin{equation}\label{eq:gu}
\gamma_U~=~\frac{3}{N\beta_T}\sum_k\frac{1}{\omega_k^2}.
\end{equation}
In the large $N$ limit, the summation can be replaced by an integral, after some arrangements, the self-consistent equation [Eq. (\ref{eq:gu})] turns into a quartic equation
\begin{equation}\label{eq:ga_u}
\gamma_U^4+4\gamma_U^3~=~\frac{9}{\beta_T^2}.
\end{equation}
The temperature dependence of $\gamma_U$ is explicit.

Predictions from the \LH{} system are quite straightforward. Eq. (\ref{eq:gamma_low}) implies that $\gamma_L$ has the following form
\begin{equation}\label{eq:ga_l}
\gamma_L~=~\frac{\langle q_n^4\rangle_{\rho}}{\langle q_n^2\rangle_{\rho}},
\end{equation}
which coincides with the prediction of the effective phonon theory (EPT) \cite{Li.06.EL}. The temperature dependence of $\gamma_L$ comes from temperature-dependent ensemble averages of the nonlinear lattice.

\subsection{Low temperature limit}
In the low temperature limit, by adopting the transfer integral operator method (TIOM) \cite{Kac.63.JMP,Scalapino.72.PRB,Krumhansl.75.PRB}, we find that(see the details in Appendix \ref{sec:tiom})
\begin{eqnarray}
\langle q^2\rangle_{\rho} &\simeq& 0.4477~T^{2/3},\label{eq:q2_lt}\\
\langle q^4\rangle_{\rho} &\simeq& 0.5458~T^{4/3},\label{eq:q4_lt}
\end{eqnarray}
which are consistent with results obtained from classical field theory \cite{Boyanovsky.04.PRD}. We then deduce that $\gamma_L$ has a following power-law temperature dependence
\begin{equation}\label{eq:gl_lt}
\gamma_L~\simeq~1.22 T^{2/3}.
\end{equation}
The validity regime of this low temperature behavior will be identified by using MD simulations together with the SST in the following subsection.

\subsection{Numerical simulations}
In this part, we utilize MD simulations to verify the above predictions. A symplectic integrator $\mathrm{SABA_2}$ with a corrector $\mathrm{SABA_2C}$ \cite{Laskar.01.CMDA} is adopted to integrate the EOMs of the $\phi^4$ lattice with a time step $h=0.02$. We take an initial condition such that the displacement of every particle is set to be zero and their velocities are randomly chosen from a Gaussian distribution at temperature $T$, after the initialization, a transient time of order $10^7$ is used
to equilibrate the system with $N=1024$.

\begin{figure}[tbh]
  \centering
  \includegraphics[width=1\columnwidth]{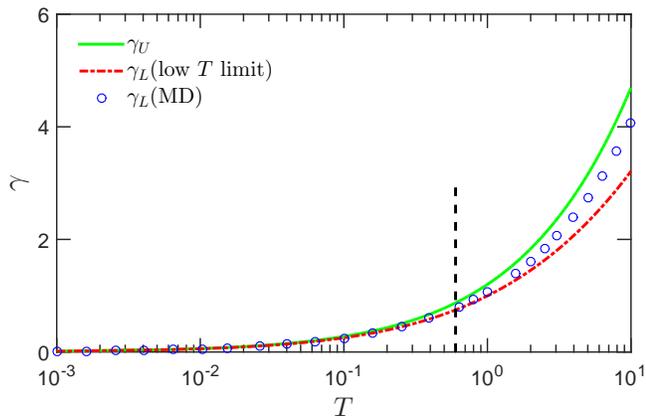}
\caption{(Color online)Temperature dependence of $\gamma$ for 1D $\phi^4$ lattice. The solid green line denotes the dependence of $\gamma_U$ [Eq. (\ref{eq:ga_u})]. The dashed-dotted red line stands for the behavior of $\gamma_L$ in the low temperature limit [Eq. (\ref{eq:gl_lt})]. The blue circles are values of $\gamma_L$ at temperature $T$ using MD results for $\langle q_n^2\rangle_{\rho}$ and $\langle q_n^4\rangle_{\rho}$ at the same temperature. The black dashed line indicates the temperature corresponds to the strong stochasticity threshold.}
\label{fig:phi4gmd}
\end{figure}

\begin{figure}[tbh]
  \centering
  \includegraphics[width=1.\columnwidth]{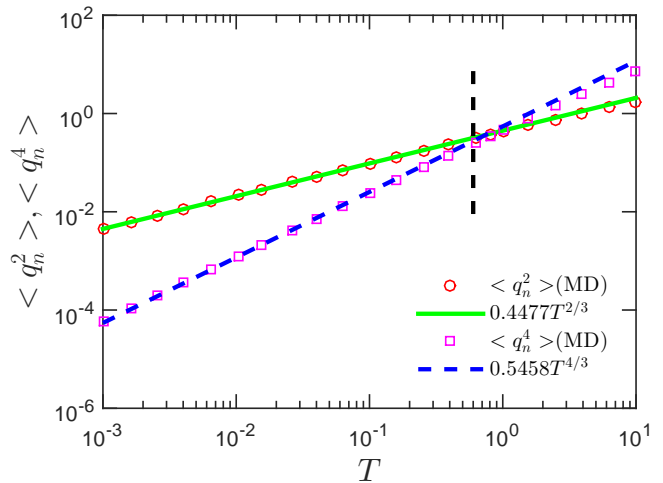}
\caption{(Color online)Ensemble averages for 1D $\phi^4$ lattice. The red circles and maroon squares are MD results for $\langle q_n^2\rangle_{\rho}$ and $\langle q_n^4\rangle_{\rho}$, respectively. The solid green line and dashed blue line denote the corresponding low temperature behaviors. The black dashed line indicates the temperature corresponds to the strong stochasticity threshold.}
\label{fig:phi4avemd}
\end{figure}

The temperature dependence of $\gamma$ are shown in Fig. \ref{fig:phi4gmd}. We determine $\gamma_U$ by numerically solving Eq. (\ref{eq:ga_u}) and getting its positive real solution. As for $\gamma_L$, we insert MD results for $\langle q_n^4\rangle_{\rho}$ and $\langle q_n^2\rangle_{\rho}$ obtained at temperature $T$ into Eq. (\ref{eq:ga_l}) to obtain its value at that temperature.
The low temperature behavior of $\gamma_L$ [Eq. \ref{eq:gl_lt}] is depicted in the figure as a dashed-dotted line. It's clearly seen that such a low temperature behavior deviates from simulation results of $\gamma_L$ when $T\gtrsim 0.6$. This is because the temperature dependence of $\langle q_n^2\rangle_{\rho}$ and $\langle q_n^4\rangle_{\rho}$ obtained from the TIOM [c.f., Eqs. (\ref{eq:q2_lt}) and (\ref{eq:q4_lt})] are no longer valid when $T\gtrsim 0.6$ compared with MD results for these two quantities as shown in Fig. \ref{fig:phi4avemd}. Noting that a temperature value $0.6$ corresponds to a roughly average energy density 0.5, which is just the SST found in the $\phi^4$ lattice \cite{Pettini.90.PRA,Pettini.91.PRA}. Therefore the so-obtained low temperature behavior of $\gamma_L$ is valid only in the weak stochasticity regime, once the system enters a strong chaotic regime, the simple power-law temperature dependence of $\langle q_n^2\rangle_{\rho}$ and $\langle q_n^4\rangle_{\rho}$ are not enough to capture the fast dynamics in the phase space.

\begin{figure}[tbh]
  \centering
  \includegraphics[width=0.9\columnwidth]{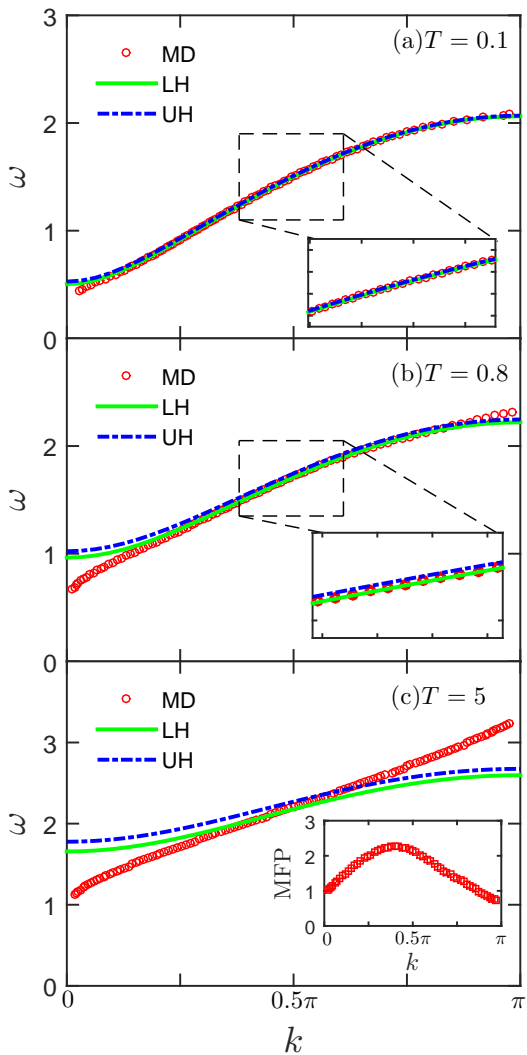}
\caption{(Color online)The dispersion for
$\phi^4$ lattice with length $N=1024$ at different temperatures. (a): a comparison between MD results (red circles) and the approximation Eq. (\ref{eq:h0d}) either with $\gamma_L$ (solid green line) or $\gamma_U$ (dashed-dotted blue line) at $T=0.1$. The inset shows details in the middle region of dispersion.
(b): a comparison between MD results (red circles) and the approximation Eq. (\ref{eq:h0d}) either with $\gamma_L$ (solid green line) or $\gamma_U$ (dashed-dotted blue line) at $T=0.8$. The inset shows details in the middle region of dispersion. (c): a comparison between MD results (red circles) and the approximation Eq. (\ref{eq:h0d}) either with $\gamma_L$ (solid green line) or $\gamma_U$ (dashed-dotted blue line) at $T=5$. The inset shows MD results for mean-free-path (MFP) of the 1D $\phi^4$ lattice at $T=5$.
}
\label{fig:phi4dis}
\end{figure}

Using the predicted $\gamma_U$ and $\gamma_L$ [c.f., Eqs. (\ref{eq:ga_u}) and (\ref{eq:ga_l})], we can check the validity of the dispersion Eq. (\ref{eq:h0d}) of optimal harmonic reference systems. We choose the tuning fork method introduced in \cite{Liu.14.PRB} to obtain MD results for actual dispersion of the 1D $\phi^4$ lattice. The comparisons between theoretical predictions and MD results are shown in Fig. \ref{fig:phi4dis}.
The figure shows results for three different temperatures. One of the temperatures ($T=0.1$) is well below the SST, implying that harmonic references should be good approximations. From Fig. \ref{fig:phi4dis}(a), we can see that this is indeed the case. The predictions of the \LH{} system and the \UH{} system both show a perfect agreement with MD results, since the values of $\gamma_L$ and $\gamma_U$ differ slightly at $T=0.1$ as depicted in Fig. \ref{fig:phi4gmd}. As the temperature increases to 0.8 which is comparable to the SST, discrepancies between theoretical predictions and MD results appear near $k=0$ and $k=\pi$ as displayed in Fig. \ref{fig:phi4dis}(b). But in the middle range of $k$, the two harmonic systems still can offer good estimations for the dispersion. The inset shows that the \LH{} system is better than the \UH{} system. When the temperature is well above the SST, the system is in a strong chaotic regime. Thus in Fig. \ref{fig:phi4dis}(c), We find for $T=5$ that our theoretical approximation Eq. (\ref{eq:h0d}) can not capture the actual dispersion of renormalized phonon in the lattice. Noting the largest MFP in this case is almost 2 as shown in the inset, it means that the phonon-phonon interaction is very strong in this regime. Such that the first-order cumulant approximation adopting in the variational principle is no longer capable of dealing with lattices in the high temperature regime, higher-order corrections to the free energy should be taken into account.

\subsection{Exact relations for ensemble averages}
More surprisingly, although the variational approach takes an approximate way to get information of nonlinear lattices, we find that it can produce some rigorous results for the nonlinear lattice. Noting the identity of the \LH{} system Eq. (\ref{eq:lh2}) and using Eq. (\ref{eq:ff}), we find $\gamma_L$ also satisfies the following equation
\begin{equation}
\gamma_L^2+4\gamma_L~=~\frac{1}{\beta_T^2\langle q_n^2\rangle_{\rho}^2}.
\end{equation}
By combining Eq. (\ref{eq:ga_l}), an identity of the 1D $\phi^4$ lattice is revealed
\begin{equation}\label{eq:equality_1}
\langle q_n^4\rangle_{\rho}^2+4\langle q_n^4\rangle_{\rho}\langle q_n^2\rangle_{\rho}~=~T^{2}.
\end{equation}
Obviously, it is distinct from the simple relation in quadratic Hamiltonians, namely, $\langle q_n^4\rangle_{\rho}=3\langle q_n^2\rangle_{\rho}^2$. Moreover, it is worthwhile to mention that this relation is valid in the whole temperature range even though $\langle q_n^2\rangle_{\rho}$ and $\langle q_n^4\rangle_{\rho}$ have different temperature dependence in the weak and strong stochasticity regime separated by the SST.

If we combine Eq. \ref{eq:equality_1} with the generalized equipartition theorem [Eq. (\ref{eq:g_equi_t})] for the $\phi^4$ lattice with $K=1, \lambda=1$
\begin{equation}
\langle 2q_n^2-2q_nq_{n-1}\rangle_{\rho}+\langle q_n^4\rangle_{\rho}=T,
\end{equation}
we will further get relations hold between various ensemble averages, for instance,
\begin{eqnarray}
&&\langle q_n^2\rangle_{\rho}^2~=~T\langle q_nq_{n-1}\rangle_{\rho}+\langle q_nq_{n-1}\rangle_{\rho}^2,\label{eq:equ_1}\\
&&\langle q_n^4\rangle_{\rho}\langle q_nq_{n-1}\rangle_{\rho}~=~\langle q_n^2-q_nq_{n-1}\rangle_{\rho}^2.\label{eq:equ_2}
\end{eqnarray}
Therefore, in this model, it is apparent that the ensemble averages of on-site quantities, e.g., $q_n^2$, can be fully determined by the ensemble averages of non-on-site terms such as $q_nq_{n-1}$.

\section{Summary}
In summary, we have extended a previously proposed variational approach to study phonon properties of general nonlinear lattices with on-site potentials, which makes our theory a possible candidate for a unified phonon theory. Intrinsic relations characterizing optimal reference system, namely, the \LH{} system and the \UH{} system, are revealed, from which we can obtain explicit forms for optimal variational parameters. Estimations for the phonon bad gap as well as the phonon dispersion are also obtained.

As a specific case, we present a thorough study of the 1D $\phi^4$ lattice by utilizing the variational approach. In the low temperature regime, a power-law temperature dependence for $\gamma_L$ qualifying the phonon band gap is found by using a complementary method, namely, the transfer integral operator method. By compared with results obtained via MD simulations, we find that such low temperature behavior is valid only in the weak stochasticity regime where the temperature is below the SST.

We further compare the theoretical predictions of the phonon dispersion with MD results for the actual phonon dispersion of the 1D $\phi^4$ lattice. At a low temperature below the SST, the agreements between theoretical and numerical results are good. At an intermediate temperature which is comparable to the SST, a discrepancy begins to appear between theoretical and numerical results. We find that the \LH{} system works better than the \UH{} system. When the system enters a high temperature regime, or equivalently, a strong chaotic regime, our theoretical predictions fails, it seems that higher order corrections to the free energy must be taken into account.

More surprisingly, we find the variational approach can produce rigorous results for the nonlinear lattice, such as an exact relation between ensemble averages of $q_n^2$ and $q_n^4$ that is valid ranging from low temperature to high temperature regime regardless of the existence of SST. This finding together with the generalized equipartition theorem further lead to interesting relations hold between ensemble averages of the 1D $\phi^4$ lattice. We then become aware of that ensemble averages of on-site quantities can be related to ones of non-on-site terms in the lattice.

\begin{acknowledgments}
The authors thank W. Si, S. Liu and N. Li for highly useful discussions. Support from the National Basic Research Program of China with Grant No. 2012CB921401 (J. Liu and C. Wu) is gratefully acknowledged. The work is also supported by the National Nature Science Foundation of China.
\end{acknowledgments}

\appendix
\section{Expressions of optimal parameters}\label{sec:opt_para}
In this part, we derive explicit expressions of optimal variational parameters for the \UH{} and \LH{} system using the identities Eqs. (\ref{eq:general_up}) and (\ref{eq:general_low}), respectively. To do this, we should utilize the virial identity \cite{Jonsson.95.JPC}
\begin{equation}\label{eq:gvi}
\left\langle \nabla\cdot\mathbf{f}\right\rangle_{\rho}~=~\beta_T\left\langle\mathbf{f} \cdot\nabla H\right\rangle_{\rho}
\end{equation}
for any polynomial $\mathbf{f}$ which depends only on coordinates. If we take a special choice $\mathbf{f}=\vec{q}$ with $\vec{q}=(q_1,q_2,\cdots,q_N)$ a vector consists of all the coordinates of the system with $N$ particles, the above equality results in \begin{equation}\label{eq:virial_theorem}
\beta_T\left\langle\vec{q}\cdot\nabla H\right\rangle_{\rho}~=~N
\end{equation}
or, specifically,
\begin{equation}\label{eq:g_equi_t}
\left\langle q_n\frac{\partial H}{\partial q_n}\right\rangle_{\rho}~=~T,
\end{equation}
which is the generalized equipartition theorem \cite{Huang.87.NULL}.

We firstly focus on the \UH{} system. Choosing $\mathbf{f}=\vec{q}(H-H_0)$ in Eq. (\ref{eq:gvi}) with the canonical measure of the \UH{} system, we get
\begin{eqnarray}\label{eq:uh1}
&&N\left\langle H-H_0\right\rangle_{\rho_0}+\left\langle\vec{q}\cdot\nabla(H-H_0)\right\rangle_{\rho_0}\nonumber\\ &&=\beta_T\left\langle\ (H-H_0)\vec{q}\cdot\nabla H_0\right\rangle_{\rho_0},
\end{eqnarray}
On the other hand, we note that Eq. (\ref{eq:virial_theorem}) holds for the \UH{} system
\begin{equation}\label{eq:uh2}
\beta_T\left\langle\vec{q}\cdot\nabla H_0\right\rangle_{\rho_0}~=~N.
\end{equation}
Substituting this equation into Eq. (\ref{eq:uh1}), we obtain
\begin{equation}\label{eq:uh3}
\left\langle\vec{q}\cdot\nabla(H-H_0)\right\rangle_{\rho_0}~=~\beta_T\left\langle\ (H-H_0); \vec{q}\cdot\nabla H_0\right\rangle_{\rho_0}.
\end{equation}
The potential of the reference system is a polynomial function of the coordinates, then the scaling operation on $H_0$ can be expressed in terms of derivatives with respect to the variational parameters $\{\chi_i\}$, i.e.,
\begin{equation}
\vec{q}\cdot\nabla H_0~=~\sum_iQ_i(\chi)\frac{\partial H_0}{\partial\chi_i}
\end{equation}
with a set of coefficients $\{Q_i(\chi)\}$ independent of coordinates, thus the right-hand-side of Eq. (\ref{eq:uh3}) vanishes due to Eq. (\ref{eq:general_up}) satisfied by the \UH{} system, we are left with
\begin{equation}
\left\langle\vec{q}\cdot\nabla H\right\rangle_{\rho_0}=\left\langle\vec{q}\cdot\nabla H_0\right\rangle_{\rho_0}.
\end{equation}
With the concrete forms of $H$ and $H_0$, the above identity gives
\begin{eqnarray}
&&\sum\limits_{s=1}f_{2s}\left\langle\delta_n^{2s}\right\rangle_{\rho_0}+\sum\limits_{k=1}g_{2k}\left\langle q_n^{2k}\right\rangle_{\rho_0}\nonumber\\
&&-\Omega_U^2\left\langle\delta_n^2\right\rangle_{\rho_0}-\gamma_U\left\langle q_n^2\right\rangle_{\rho_0}~=~0.
\end{eqnarray}
Notice the facts that if $H$ is a harmonic Hamiltonian, i.e., only $f_2$ and $g_2$ are nonvanishing in the potentials, the above equation should immediately lead to $\Omega_U^2$ and $\gamma_U$ equal $f_2$ and $g_2$, respectively, and also the nonlinearity of the on-site potential can not renormalize the  interparticle coupling strength \cite{Dauxois.93.PRE}. The above equation holds generally if and only if we let
\begin{eqnarray}
&&\sum\limits_{s=1}f_{2s}\left\langle\delta_n^{2s}\right\rangle_{\rho_0}=\Omega_U^2\left\langle\delta_n^2\right\rangle_{\rho_0},\\
&&\sum\limits_{k=1}g_{2k}\left\langle q_n^{2k}\right\rangle_{\rho_0}~=~\gamma_U\left\langle q_n^2\right\rangle_{\rho_0}.
\end{eqnarray}
These two equations lead to expressions of $\Omega_U^2$ and $\gamma_U$ [c.f., Eqs. (\ref{eq:omega_up}) and (\ref{eq:gamma_up})], respectively.

We then turn to the \LH{} system. Applying the generalized equipartition theorem [Eq. (\ref{eq:g_equi_t})] to the nonlinear system and the \LH{} system, we obtain
\begin{eqnarray}
\sum\limits_{s=1}f_{2s}\left\langle\delta_n^{2s}\right\rangle_{\rho}+\sum\limits_{k=1}g_{2k}\left\langle q_n^{2k}\right\rangle_{\rho} &=& T,\label{eq:equi_h}\\
\Omega_L^2\left\langle\delta_n^2\right\rangle_{\rho_0}+\gamma_L\left\langle q_n^2\right\rangle_{\rho_0} &=& T,\label{eq:equi_h0}
\end{eqnarray}
respectively. For the \LH{} system, the identity Eq. (\ref{eq:general_low}) with $\chi_i=\{\Omega_L^2, \gamma_L\}$ generates
\begin{eqnarray}
\left\langle\delta_n^2\right\rangle_{\rho_0} &=& \left\langle\delta_n^2\right\rangle_{\rho},\label{eq:lh2}\\
\left\langle q_n^2\right\rangle_{\rho_0} &=& \left\langle q_n^2\right\rangle_{\rho}.\label{eq:lh3}
\end{eqnarray}
Hence Eq. (\ref{eq:equi_h0}) can be rewritten as
\begin{equation}
\Omega_L^2\left\langle\delta_n^2\right\rangle_{\rho}+\gamma_L\left\langle q_n^2\right\rangle_{\rho} ~=~ T.\label{eq:equi_lh0}
\end{equation}
Subtracting Eq. (\ref{eq:equi_lh0}) from Eq. (\ref{eq:equi_h}), we get
\begin{eqnarray}
&&\sum\limits_{s=1}f_{2s}\left\langle\delta_n^{2s}\right\rangle_{\rho}-\Omega_L^2\left\langle\delta_n^2\right\rangle_{\rho}\nonumber\\
&&+\sum\limits_{k=1}g_{2k}\left\langle q_n^{2k}\right\rangle_{\rho}-\gamma_L\left\langle q_n^2\right\rangle_{\rho}~=~0.
\end{eqnarray}
Similarly, we should have
\begin{eqnarray}
&&\sum\limits_{s=1}f_{2s}\left\langle\delta_n^{2s}\right\rangle_{\rho}=\Omega_L^2\left\langle\delta_n^2\right\rangle_{\rho}\\ &&\sum\limits_{k=1}g_{2k}\left\langle q_n^{2k}\right\rangle_{\rho}=\gamma_L\left\langle q_n^2\right\rangle_{\rho}.
\end{eqnarray}
These two equations produce Eqs. (\ref{eq:omega_low}) and (\ref{eq:gamma_low}).

\section{Evaluation of ensemble averages with the TIOM}\label{sec:tiom}
In this part, we present in detail a study of the 1D $\phi^4$ lattice via the TIOM. By adopting the TIOM, we should first introduce the following transfer integral equations \cite{Riseborough.80.PRB}
\begin{eqnarray}
&&\int dq_{n-1}e^{-\beta_T K(q_n,q_{n-1})}\Psi_i(q_{n-1}) = e^{-\beta_T \varepsilon_i}\Psi_i(q_n),\label{eq:ti}\\
&&\int dq_{n}\Phi_i(q_n)e^{-\beta_T K(q_n,q_{n-1})} = e^{-\beta_T \varepsilon_i}\Phi_i(q_{n-1}),
\end{eqnarray}
where $K(q_n,q_{n-1})$ satisfies
\begin{equation}
\sum_nK(q_n,q_{n-1})~=~\sum_n\left[\frac{1}{2}\delta_n^2+\frac{1}{4}q_n^4\right],
\end{equation}
$\Psi_i$ and $\Phi_i$ are, respectively, the left-hand and right-hand normalized eigenfunctions corresponding to the eigenvalue $e^{-\beta\varepsilon_i}$. If $K$ takes a symmetric form, i.e., $K(q_n,q_{n-1})=K(q_{n-1},q_n)$, we will have $\Phi_i=\Psi^{\ast}_i$. So we choose
\begin{equation}
K(q_n,q_{n-1})~=~\frac{1}{2}\delta_n^2+\frac{1}{2}\left[\frac{1}{4}q_n^4+\frac{1}{4}q_{n-1}^4\right].
\end{equation}
The orthogonality and completeness of $\Psi_i$ are guaranteed by the Sturm-Liouville theory for Fredholm integral equations with a symmetric kernel. Using this complete set, the ensemble average $\langle g(q_n)\rangle_{\rho}$ for a local quantity $g(q_n)$ is equivalent to
\begin{equation}\label{eq:ave_lt}
\frac{\int\,\Psi_0^2(q_n)g(q_n)dq_n}{\int\,\Psi_0^2(q_n)dq_n}
\end{equation}
in the thermodynamic limit \cite{Dauxois.93.PRE,Riseborough.80.PRB,Schneider.80.PRB}, where $\Psi_0$ is the eigenfunction corresponding to the ground state. So we get
\begin{eqnarray}
\langle q^2\rangle_{\rho} &=& \frac{\int \Psi_0^2(q)q^2dq}{\int \Psi_0^2(q)dq},\label{eq:ea1}\\
\langle q^4\rangle_{\rho} &=& \frac{\int \Psi_0^2(q)q^4dq}{\int \Psi_0^2(q)dq},\label{eq:ea2}
\end{eqnarray}
where we have omitted the subscripts.

In the continuum limit and low temperatures, the transfer integral Eq. (\ref{eq:ti}) can be reduced to a pseudo-Schr\"odinger equation
\begin{eqnarray}\label{eq:sch_e}
\left(-\frac{1}{2}\frac{d^2}{dq^2}+\frac{\beta_T^2}{4}q^4\right)\Psi_i(q) &=& \beta_T^2e^{-\beta_T\varepsilon_i}\Psi_i(q)\nonumber\\
&\equiv&E_i\Psi_i(q),
\end{eqnarray}
which describes a pure quartic oscillator with a temperature dependent coefficient. Introducing a transformation $q=\lambda\tilde{q}$, the above equation becomes
\begin{equation}
\left(-\frac{1}{2}\frac{d^2}{d\tilde{q}^2}+\frac{\lambda^6\beta_T^2}{4}\tilde{q}^4\right)\Psi_i(\lambda\tilde{q})~\equiv~\lambda^2E_i\Psi_i(\lambda\tilde{q}).
\end{equation}
We choose $\lambda^6\beta_T^2=1$ or $\lambda=\beta_T^{-1/3}$ such that the transformed quartic potential is temperature independent, which implies that the transformed ground state wave-function $\Psi_0(\beta_T^{-1/3}\tilde{q})$ should also be temperature independent, thus we can denote it by $\phi_0(\tilde{q})$. Under the transformation, the averages [c.f. Eqs. (\ref{eq:ea1}) and (\ref{eq:ea2})] become
\begin{eqnarray}
\langle q^2\rangle_{\rho} &=& \beta_T^{-2/3}\frac{\int \phi_0^2(\tilde{q})\tilde{q}^2d\tilde{q}}{\int \phi_0^2(\tilde{q})d\tilde{q}}\equiv c_1\beta_T^{-2/3},\label{eq:ea1_new}\\
\langle q^4\rangle_{\rho} &=& \beta_T^{-4/3}\frac{\int \phi_0^2(\tilde{q})\tilde{q}^4d\tilde{q}}{\int \phi_0^2(\tilde{q})d\tilde{q}}\equiv c_2\beta_T^{-4/3},\label{eq:ea2_new}
\end{eqnarray}
in which $c_1$ and $c_2$ are two temperature-independent coefficients.

Since the potential $\frac{1}{4}\tilde{q}^4$ is symmetric, the ground state wave-function should be an even function of $\tilde{q}$, we only need to know the positive $\tilde{q}$ part of the function which approximately reads \cite{Liverts.06.JMP}
\begin{equation}\label{eq:psi0}
\phi_0(\tilde{q})~=~\exp\left[\int_{0}^{\tilde{q}} y(x)dx\right],
\end{equation}
where
\begin{widetext}
\begin{equation}\label{eq:yx}
y(x)~=~-\frac{\sqrt{2}}{2}x^2+\frac{2}{3}e^{z}\left[-\frac{\sqrt{2}}{2}x^2\mathrm{EI}_{1/3}(z)+x\left(\frac{3}{4}\right)^{1/3}\frac{\Gamma(2/3)}{\Gamma(1/3)}\mathrm{EI}_{2/3}(z)\right]
\end{equation}
\end{widetext}
with $z\equiv\frac{\sqrt{2}}{3}x^3$ and the exponential integral $\mathrm{EI}_{\mu}(z)$ defined by
\begin{equation}
\mathrm{EI}_{\mu}(z)\equiv\int_1^{\infty}e^{-zt}t^{-\mu}dt.
\end{equation}
Thus we can determine $c_1$ and $c_2$ by numerically calculating the integrals with the above form of $y(x)$, it turns out that the $c_1$ and $c_2$ equal $0.4477$ and $0.5458$, respectively. Therefore, we get the forms Eqs. (\ref{eq:q2_lt}) and (\ref{eq:q4_lt}) in our analysis.

%\bibliography{a}

\end{document}